**Synthesis of Methotrexate loaded Cerium fluoride nanoparticles with pH sensitive extended release coupled with Hyaluronic acid receptor with plausible theranostic capabilities for preclinical safety studies**

# MASTER THESIS

Submitted by

**Nitish Manu George**

In partial fulfilment for the award of the Degree of

**BACHELOR & MASTER OF TECHNOLOGY (DUAL)**

**IN**

**NANOTECHNOLOGY**

Under the guidance of

**Dr. Mayuri Gandhi**

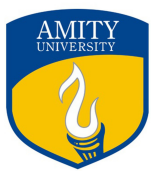  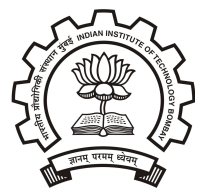

**Amity University, Noida**           **Indian Institute of Technology-Bombay**

**Uttar Pradesh, India**                **Mumbai, India**

**December – May, 2016**

# Acknowledgements

I'd like to extend my sincerest gratitude towards Dr. Mayuri Gandhi for accepting my request to work under her guidance at the C.R.N.T.S. facility in I.I.T. Bombay. Her continued support, timely interventions and sound advice throughout the course of my project work is what lead to its eventual success. Although her nature of expecting perfection would often push me to continually improve and hone my skills, her friendly demeanour made her a guide who was easy to approach and connect with on a personal level. The warmth she showered upon me certainly instils in me a feeling that I wasn't wrong in pursuing my research under her experienced guidance.

I'd also like to thank all those working in the facility for being warm and friendly. On more than one occasion, they've extended their sincerest assistance in my project. Without some key recommendations, I wonder whether I'd even be able to complete my project. It was a humbling experience to see them take time out from their schedules in order to assist me with various challenges I faced. From constant inquiries towards the status of my project to subtle words of encouragement, the contribution of the laboratory members towards my project can't be overlooked. More often than sometimes, I definitely would've been an annoyance to them but they never held it against me and chose to only support me to the best of their capabilities. They are all a true inspiration in professionalism with a hint of familial bonding in the workplace.

Furthermore, I'd like to thank all the researchers who've worked previously in the subject enabling me to develop my own hypothesis by being inspired from their work. Among them, my own friend Varun S., who I'd like to acknowledge as an informed researcher who'd assist me the best he could since my project directly takes a page from his previous research. He'd been generous enough to share all forms of relevant information in the spirit of goodwill towards research and on more than one occasion encouraged me.

Lastly, I'd like to thank my internal guide Dr. Nidhi Chaunhan and all the faculty members of Amity Institute of Nanotechnology at Amity University for their support and guidance. I'd also like to extend my gratitude towards my parents who supported my financially and emotionally even though there was a familial crisis at hand. I'd also like to thank my friends who helped in keeping me in high spirits throughout the course of my project.

**Abstract**

A key challenge in drug delivery systems is the real time monitoring of delivered drug and subsequent response. Recent advancement in nanotechnology has enabled the design and preclinical implementation of novel drug delivery systems (DDS) with theranostic abilities. Herein, fluorescent cerium fluoride ($CeF_3$) nanoparticles (nps) were synthesized and their surface modified with a coat of polyethylenimine (PEI). Thereafter, Methotrexate was conjugated upon it through glutaraldehyde cross-linking for a pH-sensitive release. This was followed by the addition of a Hyaluronic acid (HA) receptor via 1-Ethyl-3-(3-dimethylaminopropyl)-carbodiimide and N-hydroxysuccinimide (EDC-NHS) chemistry to achieve a possible active drug targeting system. The obtained drug delivery nano-agent retains and exhibits unique photo-luminescent properties attributed to the nps while exhibiting potential theranostic capabilities.

# Contents

# List of Figures

# 1. Introduction

## 1.1. Cancer Research

Cancer is one of the most popular diseases on the face of this earth that has existed for almost the entirety of human history. With over 200 known variants, it is a group of diseases involving abnormal growth of cells with the potential plausibility of invading or spreading to healthier cells or parts of the body. Lacking in specific fingerprints and having the widest spectra of causalities, it is next to impossible to prove the cause for a particular cancer regardless of its type.[1] Therefore, a lot of research is under way to identify the causes of the disease as well as in developing strategies for its prevention, diagnosis, treatment and subsequent cure.

Although a number of treatment options exist for cancers such as surgery, chemotherapy, radiation therapy, gene therapy, immunotherapy, photo-dynamic therapy, targeted therapy, palliative care etc.,[2-13] they still do not accommodate for the different phenotypes observed in individual cases or provide a generic therapeutic to be followed. Also, cancer is a class of disease, it can also be argued that it will be be highly unlikely that a single cure for cancer will ever come to exist. However, research continues in hopes of achieving a "silver bullet" treatment technique that can be employed over a majority of the cases.

Due to the dynamic nature of the disease, Cancer Research is one of the most funded forms of medical research in the world with billions spent worldwide every year. Recent advancements in molecular and cellular biology has lead to a revolution bringing about better treatments through a variety of preclinical and clinical trials reducing the risk quotient involved with the treatment of the disease with every passing day.[14] Perhaps one day, the silver bullet treatment that every researcher working in the field envisions might even become a reality with all the awareness, effort and investment being put into the field.

## 1.2. Drug Delivery Systems for Cancer

There exists a multitude of therapeutic techniques for cancer but drug delivery remains one of the most prominent. This is due to the fact that drugs may be used to limit or inhibit the spread of the disease as well as sometimes used as a measure to cure it. Depending upon individual requirements drug delivery systems tend to involve either deliverance of the drug directly into the affected area or to an affected area through utilization of a carrier. With advancements in technology, the carrier techniques have been offered additional functionalities such as specific molecular targeting enabling higher efficiencies. Current efforts include development of novel DDS that activate only in the target area with sustained or controllable release.[15-22]

## 1.3. Theranostics

Theranostics is the field of science pertaining to the combination of diagnosis and therapeutics. It involves utilization of agents of drug delivery that are capable of carrying drugs to their target locations while offering limited but crucial diagnostic information of the environment around them in real-time. Recently, biomedical research pertaining to DDS have come to realize the immense significance of controlled targeted release of drugs along with the ability to monitor drug effect and the patient response in real-time. Researchers now realize that the characteristics shall enable physicians to make changes in therapeutic techniques employed based upon feedback received, while simultaneously, reducing damage to healthy cells and organs from drug side-effects. However, such theranostic capabilities are yet largely limited to in-vitro environments and preclinical studies due to the extraordinarily enormous amount of research required to get such an agent to be safe and approved for human use. Nevertheless, the contribution of the same within its current limited domain is not to be overlooked as it still enables one to gather more information regarding individual phenotypes exhibited by patients and their possible response to a therapeutic method. [23-33]

### 1.4. pH sensitive glutaraldehyde cross-linking

Glutaraldehyde is one of the most versatile cross-linking reagents that is prominently utilized in biochemistry. Present in at least thirteen different forms based on solution conditions, there is substantial literature available regarding the use of the reagent for various cross-linking applications with various carbohydrates, lipids, nucleic acids, enzymes, and other soluble and insoluble molecules. Owing to many discrepancies and the unique characteristics of each moiety linked with the reagent however, cross-linking procedures are largely developed through empirical observations and its subsequent properties are also tested in the same manner. However, many research articles elucidate cases wherein the cross-linking acts in a pH sensitive manner with respect to linking efficiencies and reversibility of the link formed. [33-35]

### 1.5. Hyaluronic Acid – Cancer receptor

Hyaluronic acid is a unique nonsulfated glycosaminoglycan which is formed in the plasma membrane instead of the Golgi. Through literature it has been mentioned that the macromolecule significantly contributes to cell proliferation and migration and has also been linked to the progression of many malignant tumours. For cancer, HA acts as a receptor enabling targeting due to the over-expression of CD44 glycoprotein in cancer cells which is recognized as a major receptor of HA.[35-44]

### 1.6. Objective

Through this thesis, an attempt at creating a novel DDS for Cancer, utilizing $CeF_3$ nps is described. The naked lanthanide nps have previously been associated with providing protection against oxidative stress as well as being toxic. The host material was selected under the assumption that the same characteristic shall be reproduced in preclinical safety studies even when functionalized as a drug delivery agent. If realized, it would entail the safeguard of healthy cells from apoptosis due to oxidative stress from introduction of nanoparticles into their environment. Furthermore, the understanding that $CeF_3$ can act as a scintillator makes it all the more desirable as a drug carrier for the purpose of theranostic applications; at least within the scope of preclinical studies.

Utilizing PEI, the surface of the nanoparticles synthesized were modified such that glutaraldehyde cross-linking of the amine groups in PEI with the amine group in the Cancer drug – Methotrexate - was enabled. This entailed that the conjugation of the drug was pH-sensitive which consequently will allow controlled release within Cancer cells alone. This would eventually reduce chances of damaging healthier cells rendering the drug more effective than when delivered using conventional means.

In order to impart a targeting functionality to the DDS, Hyaluronic acid was added as a receptor. It is well established in literature as a targeting compound for Cancer cells and it was implemented to the system via EDC-NHS coupling. With the addition of the final component, the novel DDS that was planned was completed. With unique optical properties of $CeF_3$ nps coupled with the enhancements, it is believed that the DDS may prove beneficial in theranostic applications across preclinical screens; with significant possibility of the same to be transferable to in-vivo conditions upon optimization.

## 2. Materials and Methods

### 2.1. Reagents

Ammonium fluoride ($NH_4F$), Cerium nitrate hexahydrate ($Ce(NO_3)_3 \cdot 6H_2O$), Polyethylenimine (($C_2H_5N)_n$), Methotrexate ($C_{20}H_{22}N_8O_5$), 25% Glutaraldehyde solution ($C_5H_8O_2$), Hyaluronic acid sodium salt ($C_{14}H_{22}NNaO_{11}$), Sodium chloride (NaCl), Potassium chloride (KCl), Disodium phosphate ($Na_2HPO_4$), Monopotassium phosphate ($KH_2PO_4$), 35% Hydrochloric acid, 1-Ethyl-3-(3-dimethylaminopropyl)-carbodiimide (EDC), N-hydroxysuccinimide (NHS) were all purchased from Sigma Aldrich. Absolute Ethanol was supplied by Changshu Hongsheng Fine Chemical Co., Ltd. Deionized (DI) water was procured from a pre-existing distillation unit within the department.

### 2.2. Experiment

**a) Preparation of Phosphate Buffered Saline (PBS) solution**

1 L PBS solution with pH 7.4 was prepared as per the recipe elucidated in Cold Spring Harbor Protocols. 250 mL of which was later on adjusted to pH 5.0.[45]

**b) Synthesis of Cerium fluoride nps**

Cerium fluoride nanoparticles were synthesized following a slightly modified version of the simple co-precipitation route employed by Varun et. al.[46] wherein 40 mM cerium nitrate hexahydrate dissolved in ethanol was mixed with hot (~70º C) 120 mM ammonium fluoride hexahydrate dissolved in DI water and stirred continuously for 10-15 minutes.

**c) Surface Modification and Drug Loading**

To the freshly prepared solution with cerium fluoride nps precipitation, 1% PEI in DI water was made and added in a 1:25 ratio by solvent. It was then left on stirring for nearly 24 hours in order to coat the nps with the polymer coating. 1 mM of the drug was then added to the solution and left to stir for another hour for a uniform dispersion. This was followed by the addition of nearly 1.5 times absolute ethanol to the amount of solvent present in the solution in order to attain desolvation. After another hour of stirring, a few drops of 25% glutaraldehyde solution was added and the solution was left to stir for about 12 hours. The resulting solution was made to precipitate through heating and addition of ethanol (if required) and thereafter washed with ethanol. It was then re-dispersed in DI water/PBS as per requirement.

### d)   Preparation and coupling of Hyaluronic acid

A 10 mM solution of Hyaluronic Acid Sodium Salt was prepared in 10 mM solution of HCl. Thereafter 10 mM of EDC and NHS were added to the solution and stirred for about 5 minutes before adding immediately to the drug-PEI-nps conjugate dispersion prepared earlier and left for stirring for nearly 24 hours. The final product was thereafter washed with hot DI water. It was then re-dispersed in DI water/PBS as per the requirement.

### e)   Drug release studies

1 mM of the drug was dispersed in 10 mL of pH 7.4 PBS solution and 300 uL of it was further diluted in 1.2 mL of pH 7.4 PBS solution. This was then characterized using UV-Vis spectroscopy to ascertain maximum peak of the drug. The supernatant post washings were collected and air dried. It was then re-dispersed in 10 mL of pH 7.4 PBS and then further diluted by taking 300 ul in 1.2 mL of pH 7.4 PBS before characterization using UV-Vis to evaluate drug loss during drug loading and receptor attachment. Thereafter, a certain weight of the prepared agent was dispersed in 10 mL of pH 7.4 PBS and loaded into a dialysis membrane. This was suspended in 50 mL solution of pH 7.4 PBS for nearly a day in order to remove any drug releasing at that pH. Immediately after this, the agent loaded membrane was suspended in 150 mL of pH 5.0 PBS solution. 300 uL samples were taken out on a time lapse and the removed PBS was replenished in order to carry out drug release studies by creating dilutions in 1.2 mL of pH 5.0 PBS solution and then characterizing using UV-Vis spectroscopy. The same weight of the drug was directly dispersed in 10 mL of pH 5.0 PBS and characterized for drug max value.

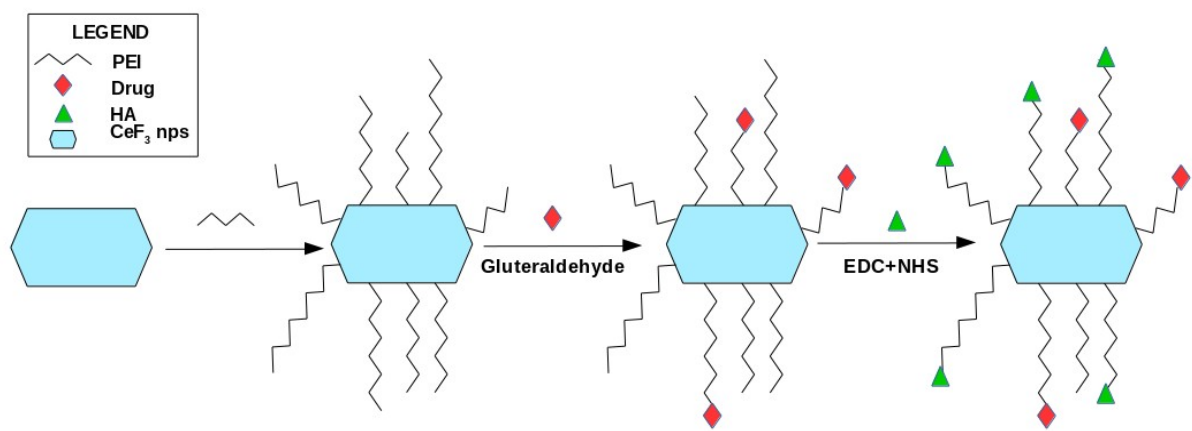

*Figure 1: Schematic representation of proposed agent preparation.*

## 2.3. Characterization

UV-Vis absorption spectra were acquired by using the Cary 100 UV-Vis Spectrophotometer by Agilent Technologies. X-ray powder Diffraction (XRD) Spectroscopy was carried out on a Philips powder diffractometer PW 3040/60 X'Pert Pro with Cu Kα radiation to determine nps characteristics. Photo-luminescence (PL) for the emission spectra of the sample was carried out on a Horiba Jobin-Yvon FluroLog-3 model. A Bruker Hyperion 3000 Microscope with a Vertex 80 Fourier transform infra-red (FTIR) system was used for obtaining FTIR spectra. The data was then plotted using the open source software GNU-PLOT.

## 3. Results

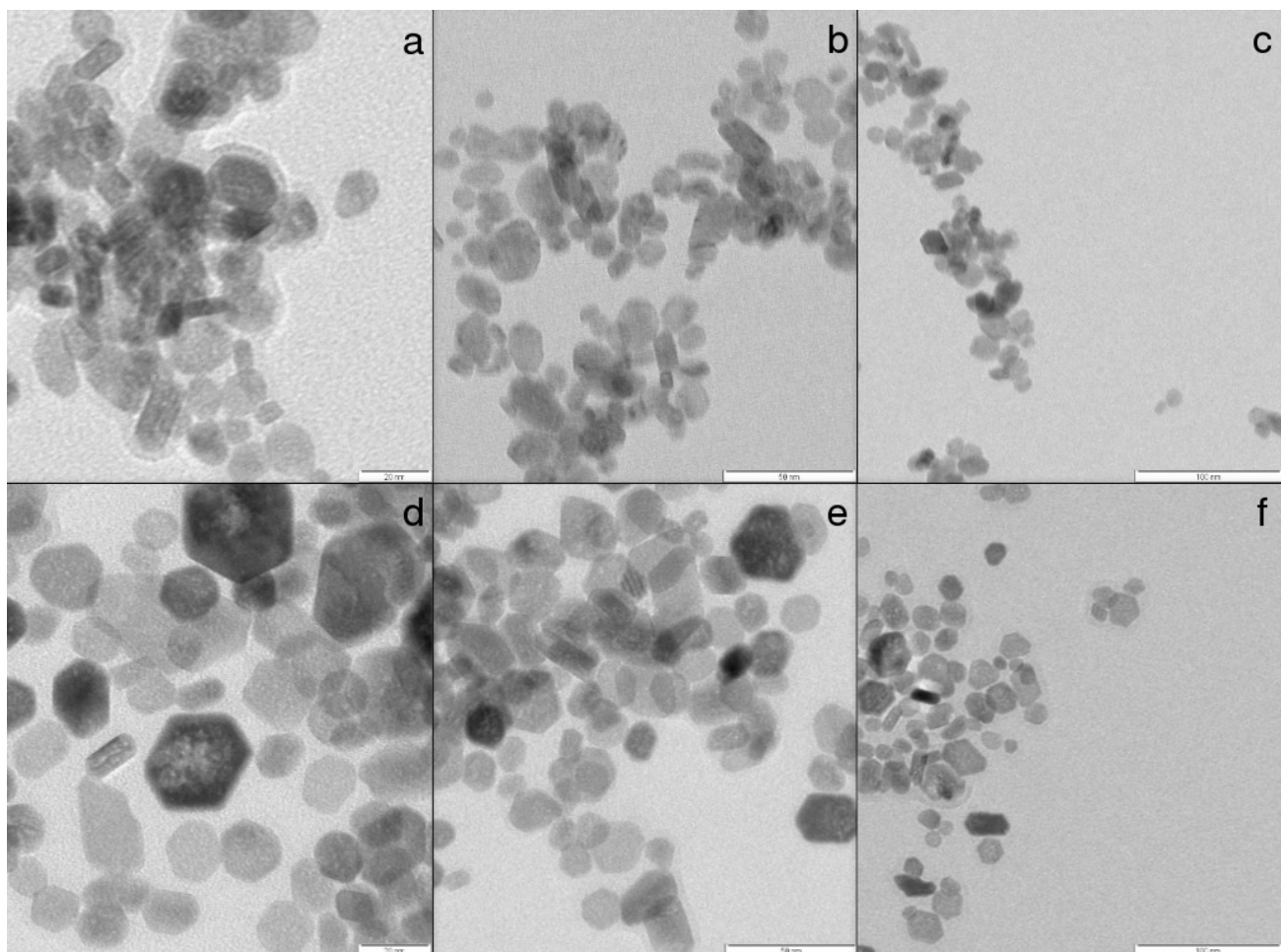

*Figure 2: TEM image. a) Drug-PEI-nps without HA, scale: 20 nm b) Drug-PEI-nps without HA, scale: 50 nm c) Drug-PEI-nps without HA, scale: 100 nm d) Drug-PEI-nps with HA, scale: 20 nm e) Drug-PEI-nps with HA, scale: 50 nm f) Drug-PEI-nps with HA, scale: 100 nm.*

From the image, it is clear that cerium fluoride nanoparticle were formed with an average size range of 10-20 nms. Upon close inspection, translucent enclosures can be seen around the particles which might be the PEI layer functionalized with the drug and receptor. Upon long periods of exposure, it was observed that this layer would turn invisble and therefore remained mostly unobservable. The hypothesis for the phenomenon is that the organic PEI layer might be getting burned off owing to the high intensity electron beam of the TEM. It was the major reason why images of CeF3 nanoparticle with PEI coating was not included since the images would go blank within seconds of exposure. With the functionalization of the organic layer however, the same phenomenon was limited, perhaps due to the quenching of the light sensitive PEI.

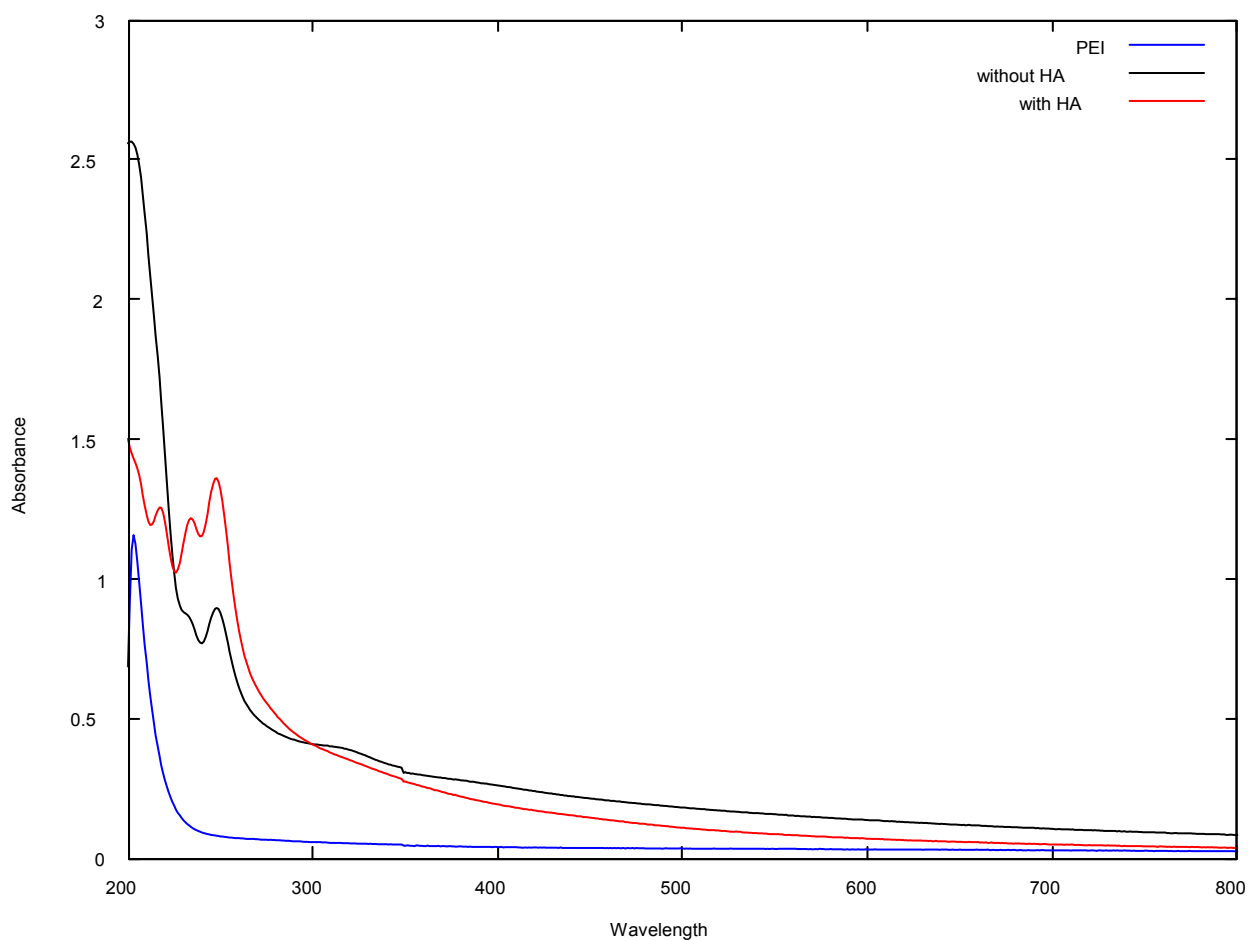

*Figure 3: Graph plot showing data acquired from UV-Vis Spectroscopy for prepared Samples. (Units: Absorbance – a.u.; Wavelength - nm)*

The graph above shows excitation peak is in line with $CeF_3$ nanoparticles i.e., close to 250 nm and shows a rise in absorbency with the addition of HA receptor. This entails that the agent would absorb more light resulting in a weaker emission spectra of the nanoparticle core due to the presence of larger molecules over the surface. However, the agent does retain characteristics of its core which is evident from an almost similar UV-Vis pattern.

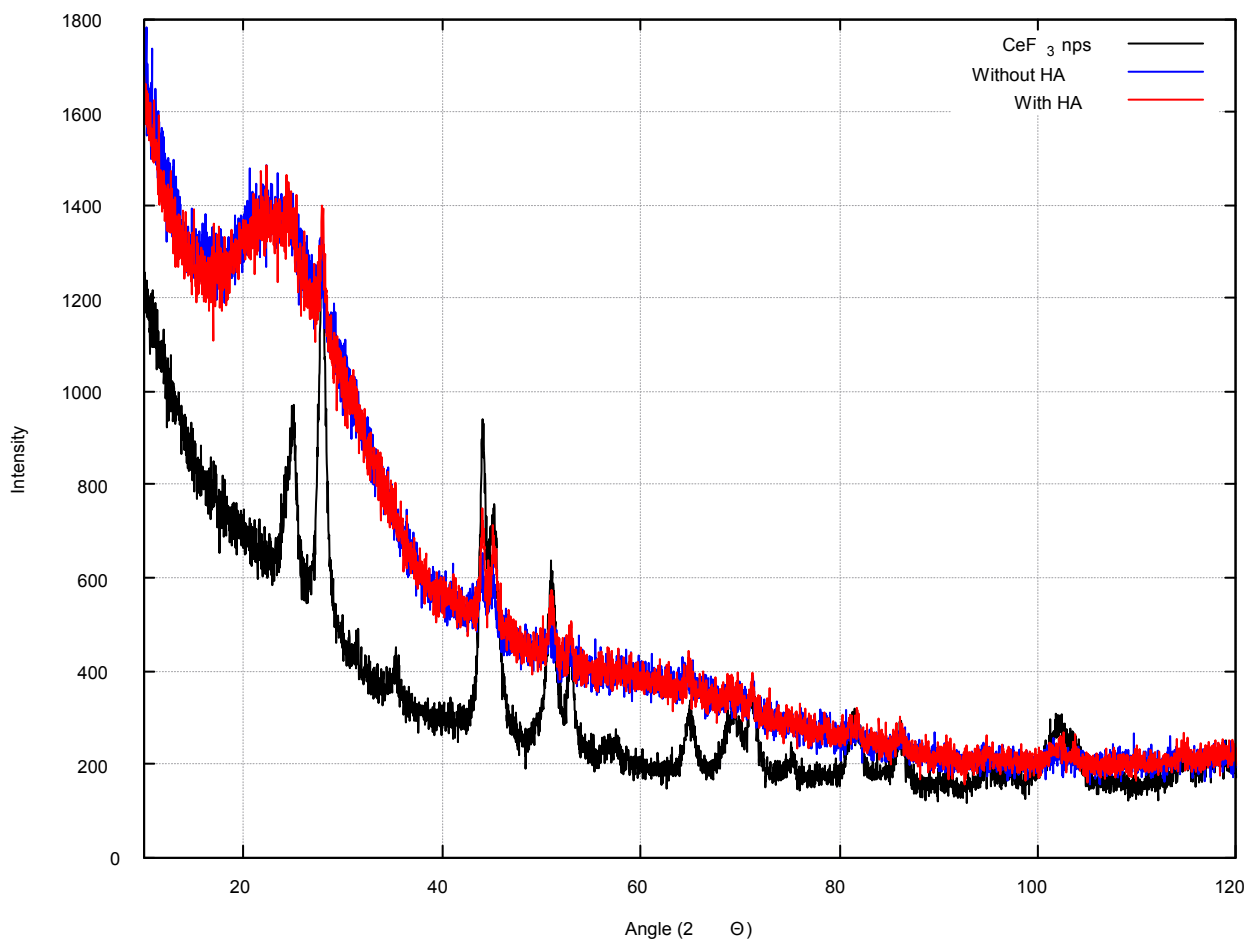

*Figure 4: XRD plot of prepared nanoparticle samples in native as well as conjugated form.*

The XRD data of regular naked $CeF_3$ nps received coincided with the XRD peaks found in literature. However, when in conjugated form, the peaks were rather indiscernible owing to the high level of noise from the organic polymer attached. It can also be argued that the crystalline structure is no longer prominent in the agent which results in such a trend in the XRD graph above.

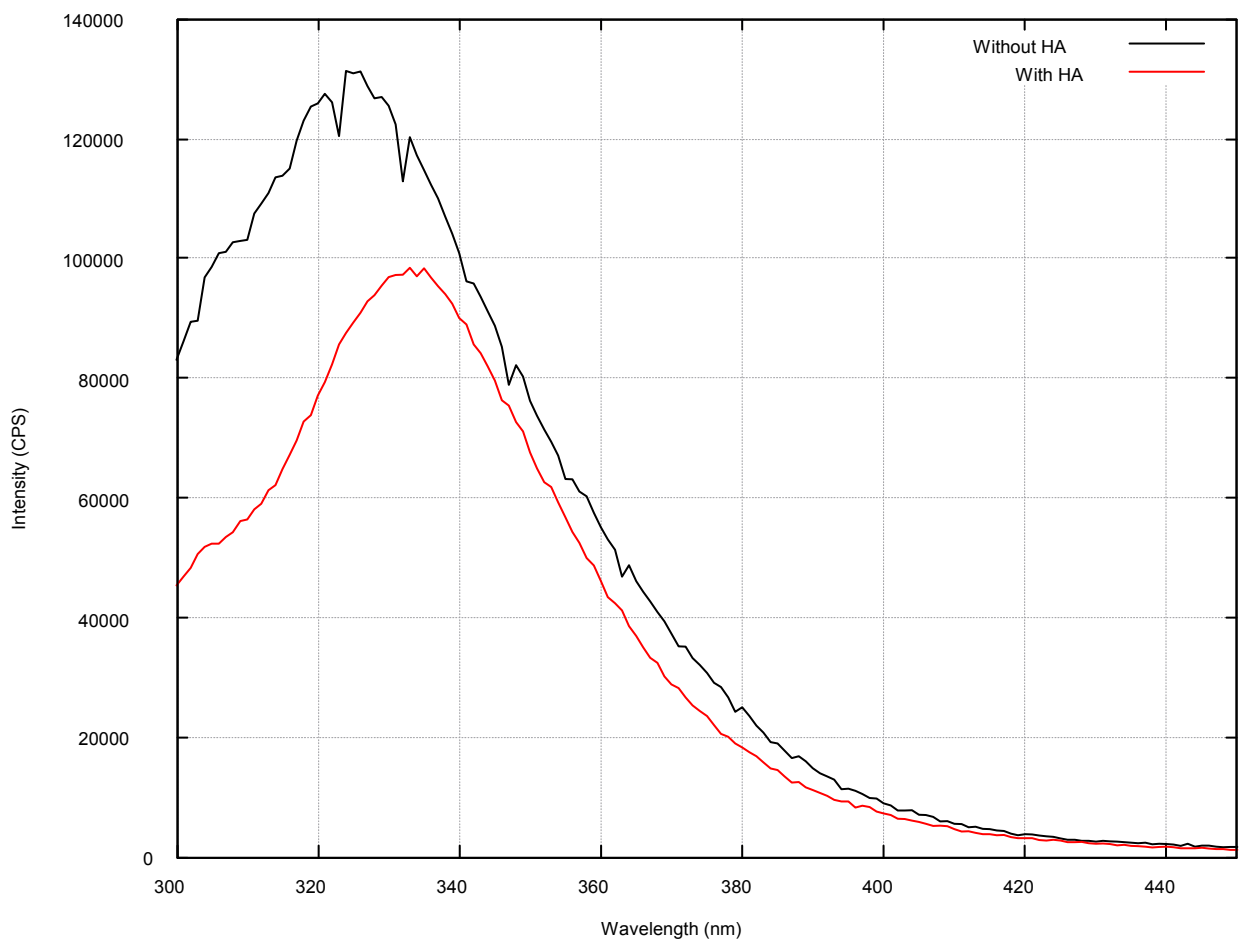

*Figure 5: PL of samples without and with HA.*

The PL emission spectra of the resulting agent was coherent with the PL spectra of $CeF_3$ nps in literature. Although there was a significant reduction in the emission intensity, it put to rest concerns of the emission being completely quenched and resulting in the agent being useless for theranostic purposes. In order to enhance the emission of the agent, the core nps system may be doped with impurities such as Europium or Terbium which are almost the same size and are in all probability non-toxic in nature.

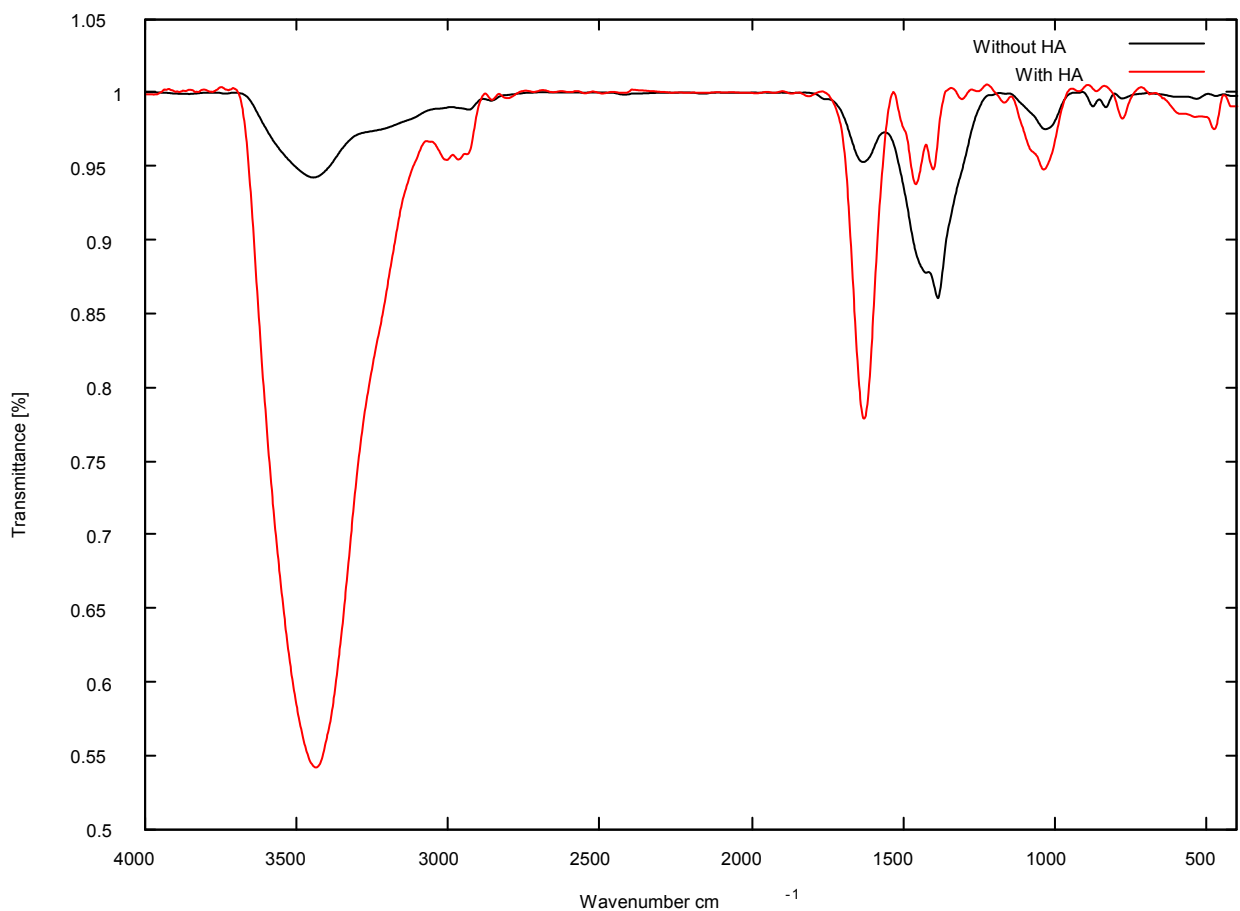

*Figure 6: FTIR plot of prepared samples without and with HA.*

FTIR plot comparing samples with and without HA shows an increase in the 3500-3300 nm region post addition which could be due to the increase of O-H and N-H bonds due to the addition of the glycosaminoglycan. The amine and alcohol or phenol stretch is quite significant as inferred from the plot above. An increase in the amide C=O stretch is also observed. There was a significant decrease in the peak in the 1300-1400 nm region however, the trend is pretty much similar. Since the region close to 1200 nm is known as the fingerprint region, it can be hypothesised that although a lot of C-O bonds might have been lost in the process of HA addition, the agent retained the drug. For further clarification, a standard FTIR spectra of naked $CeF_3$ nps and one with PEI attached could be taken alongside and compared.

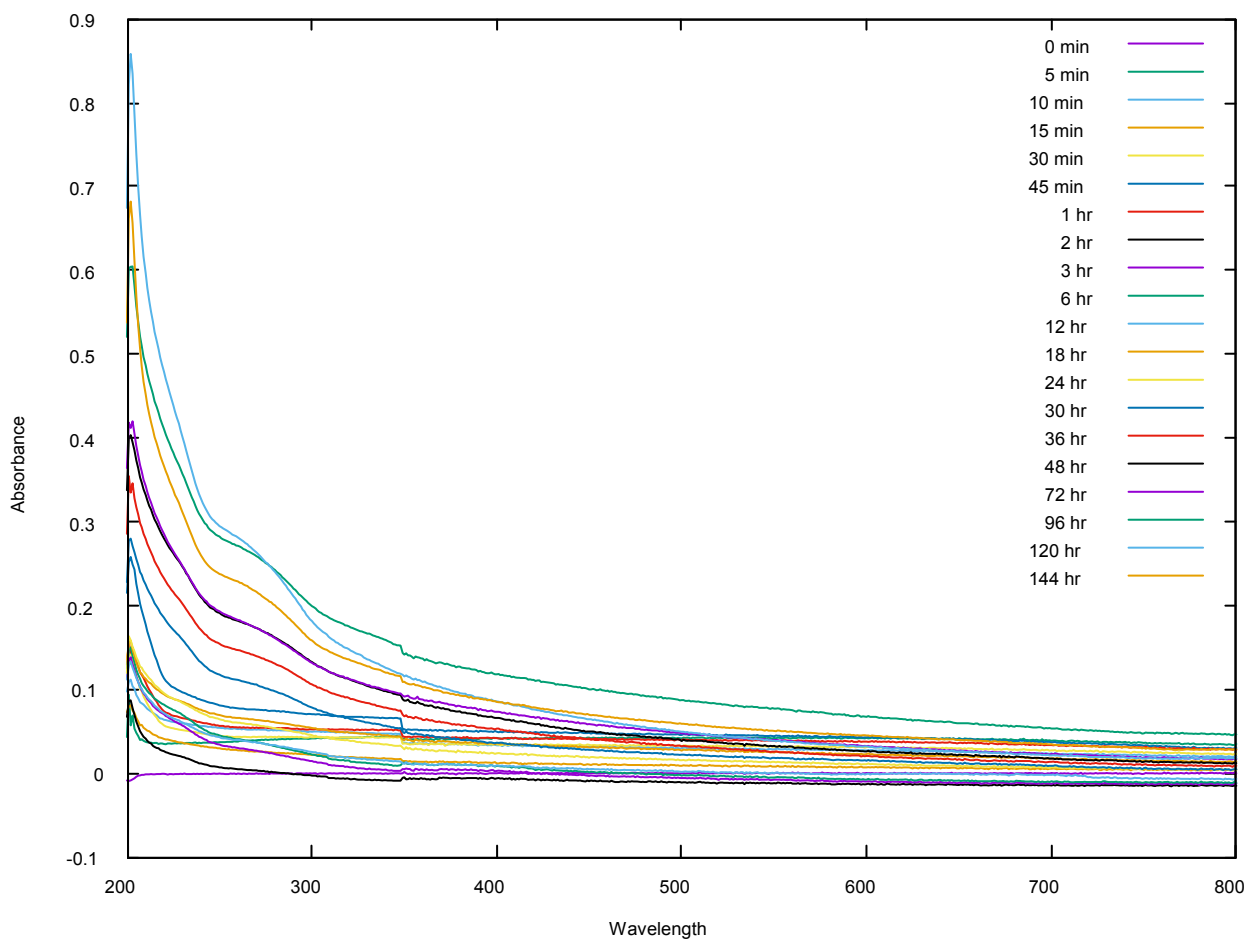

*Figure 7: Drug release studies over time-lapse. (Units: Absorbance – a.u.; Wavelength - nm)*

Drug release studies were conducted over a time lapse of – 5 minutes, 10 minutes, 15 minutes, 30 minutes, 1 hour, 2 hours, 3 hours, 6 hours, 12 hours, 18 hours, 24 hours, 30 hours, 36 hours, 48 hours, 72 hours, 96 hours, 120 hours and 140 hours. An extended release was not expected. However, since it was observed upon characterization. Fluctuations in drug release data might be a result of a combination of human error, instrumental error as well as re-absorption of the drug by the agent.

## Discussions and Conclusions

Although the agent offers promising results in terms of drug release as well as photoluminescence, and core being lanthanide would offer obstruction to X-rays offering another method of agent monitoring within the body, it is yet to be tested for toxicity. The behaviour of the prepared agent is but anyone's guess unless a few toxicity assays are performed to determine their safety. Nevertheless, the system is proposed to be useful within the scope of preclinical safety studies wherein cultured cells are utilized. In this manner, newly formulated drugs can be tested in a safe manner without animal or human casualties. Further tests are required to conclude the targeting capabilities owing to HA.

Due to the dynamic and versatile nature of the agent, it can be highly tuned to suit specific needs. The luminescence can be enhanced via means of introducing impurities of other lanthanide elements. Also, the application method can be varied. Addition of an external encapsulation could enable drug delivery in an oral form. Transferring the agent onto a thin film can also help in facilitating topical approaches to certain malignant tumours on the skin surface. Since nanoparticle systems have a high penetrative capability, it can be a good way to combat skin cancer in many individuals and also in a very cheap and easy way.

In conclusion, a novel DDS was designed utilizing nps from the lanthanide series as the core element. The results observed seem rather promising towards evolving the current theranostic techniques available and might prove crucial in the creation of a marketable technology. However, an immense amount of research would be required prior to technology transfer including a number of toxicological screens, preclinical and clinical trials along with a number of approvals. Nevertheless, if realized, the technology developed shall revolutionize modern medicine.